\documentstyle[aps,prd,eqsecnum,epsf]{revtex}

\newcommand{\square}{\kern1pt\vbox{\hrule height 1.2pt\hbox{\vrule width
1.2pt\hskip 3pt\vbox{\vskip 6pt}\hskip 3pt\vrule width 0.6pt}\hrule height
0.6pt}\kern1pt}
\newcommand{\beq}{\begin{equation}}
\newcommand{\beqn}{\begin{eqnarray}}
\newcommand{\eeq}{\end{equation}}
\newcommand{\eeqn}{\end{eqnarray}}

\begin{document}

\draft


\title{Covariant Gravitational Equations on
  Brane World with Gauss-Bonnet term}

\author{Kei-ichi Maeda$^{1,2,3}$, and Takashi Torii$^{2}$}

\address{$^1$Department of Physics, Waseda University,
Shinjuku,  Tokyo 169-8555, Japan\\[-1em]~}
\address{$^2$ Advanced Research Institute for Science and Engineering,
Waseda  University, Shinjuku, Tokyo 169-8555, Japan\\[-1em]~}
\address{$^3$ Waseda Institute for Astrophysics, Waseda University,
Shinjuku,  Tokyo 169-8555, Japan\\[-1em]~}

\date{\today}

\maketitle

\begin{abstract}
We present the covariant gravitational equations to describe a
four-dimensional brane world
in the case with the
Gauss-Bonnet term in a bulk spacetime, assuming  that  gravity is confined 
on the
$Z_2$ symmetric brane.
It contains some components of five-dimensional Weyl curvature 
($E_{\mu\nu}$)  which
describes  all effects from the bulk spacetime
just as in the case of the Randall-Sundrum second model.
Applying this formalism to cosmology,
we derive the generalized Friedmann equation and
calculate the Weyl curvature term, which is directly obtained from a black
hole solution.
\end{abstract}

\section{Introduction}
A brane is now one of the most important
ideas in particle physics\cite{brane}.
It may provide us  a new solution for the so-called hierarchy problem and a
new mechanism for compactification of extra dimensions.
  Since the fundamental scale could be
TeV in some models\cite{Arkani-Hamed}, a gravitational effect
is not ignored  even at  much lower energy scale than the Planck
mass. For example, a black hole formation in the next generation
particle collider could be observed\cite{BH-formation}. It should
be also stressed that we could come across the first experimental evidence of
quantum gravity.
It may  also change our view of the universe: we live in a 4-dimensional
(4-D)
hypersurface embedded in a higher-dimensional bulk spacetime\cite{early}.
By these reasons, the brane world scenarios
attract many attention.

Among many brane models, ones  proposed by Randall and
Sundrum are very important\cite{Randall_Sundrum1,Randall_Sundrum2}.
They are motivated
by superstring/M-theory, i.e., the orbifold compactification of
higher-dimensional string theory by the dimensional reduction of
11-dimensional supergravity in $R^{10}\times S^1/Z_2$\cite{Horava}. The
standard-model particles are confined in a 4-D brane world
while gravity accesses extra dimensions like a
string/D-brane system.
  In their first model (RS I)\cite{Randall_Sundrum1}, they proposed a mechanism
to solve the hierarchy problem with two branes, whereas in their second model
(RS II)\cite{Randall_Sundrum2}, they considered a single brane  with a
positive tension,  where 4-D Newtonian gravity is recovered at low energies
even if the extra dimension is not compact. This mechanism provides us an
alternative compactification of extra dimensions.
However, those models may inevitably  expect a singular spacetime just as
in general relativity, although they are based on a string theory.
In fact,
Maldacena and Nunez  showed no-go
theorem\cite{Maldacena}, which states that there are no non-singular
warped compactifications in a large class of supergravity theories
including 11-dimensional supergravity, IIB, IIA and massive IIA. One
of the ways to evade this argument is adding the higher curvature
corrections to the bulk Lagrangian. The higher curvature terms
naturally arise as a next leading order of the $\alpha'$-expansion of
a superstring theory\cite{Gross}. One may expect that they are described by
the so-called Gauss-Bonnet
combination, which  is shown to be a ghost-free combination\cite{Zwie}. It
also plays a fundamental role in Chern-Simon theories\cite{Chamseddine}.
It was  shown that the graviton zero mode is localized at law energies  in
the Gauss-Bonnet brane system as in the RS II model\cite{Kim,Meissner,Neupane}
and that the  correction of the Newton's law becomes milder by including the
Gauss-Bonnet term
\cite{GB_brane_Newton}.

As for cosmology with a brane world,  there
has been a lot of works  over the last several
years\cite{brane_cosmology1,brane_cosmology2,brane_cosmology3}.  In
particular, based on the RS II brane model, which is one of most popular
ones,  some interesting properties such as ``dark radiation" or
quadratic density term in the Friedmann equation have
been found, assuming  a simple  bulk
metric\cite{brane_cosmology3}.
Since gravity is confined on a brane, the induced metric describes
gravity on a brane. Hence the geometrical reduction gives
a covariant form of the basic equations for brane 
gravity\cite{SMS,MW,MMT,M_supple}.
Applying this formalism, we find the  Friedmann equation easily.
 
As we discussed above, since including
the Gauss-Bonnet term is important, such models are also  extensively
studied\cite{Cline,Deruelle,Binetruy,Lidsey,Luty,GB_brane_holography}.
Many authors so far studied mainly in the contexts of a resolution of initial
singularity, inflation and a self-tuning mechanism of cosmological constant.
In  these analysis, a simple  bulk  metric is assumed just as in
Ref.~\cite{brane_cosmology3}.

In order to understand those problems further, it
may be convenient for us to extend the covariant
gravitational equations on a brane to the case with the Gauss-Bonnet term.
This is the purpose of the present paper.
To find such equations, we first have to
prove a consistency with a thin-wall ansatz.
When we have a  system with quadratic  curvature terms in a bulk spacetime, we
will be soon faced with an obstacle. In general, we expect terms such as
$(\pounds_n K_{AB}$$)^2$ in the field equations, where $K_{AB}$ is the
extrinsic curvature of a brane.  If a brane is an infinitely thin singular
wall, which could be described by
the $\delta$-function, the extrinsic curvature must have a jump at a brane.
However, if $\pounds_n K_{AB}$ is proportional to the $\delta$-function,
a term of $(\pounds_n K_{AB}$$)^2$ makes troubles because it gives
a  square of the
$\delta$-function.
The reason for this breakdown is  our thin-wall ansatz.
We  have to find other
relevant junction condition which may require information about an internal
structure of a brane,
that is, we have to discuss a thick brane model.
  The basic equations may not be
described only by a geometric reduction.

In the case with the Gauss-Bonnet term, however, the situation changes
completely.
The basic equations  show  a quasi-linear property
pointed out by Deruelle and Madore\cite{GB_brane_linear}, which guarantees
a thin-wall ansatz because it contains only linear terms of  $\pounds_n
K_{AB}$. Using this fact, some authors derived the generalized Friedmann 
equation
with  a simple  bulk  metric.
With this fact, here we derive the covariant gravitational
equations on a brane in the case with the Gauss-Bonnet term.
The basic equations are described by 4-D brane variables except for
the 5-dimensional (5-D) Weyl curvature tensor $E_{\mu\nu}$.
  Although our system is not
closed because of  the existence of  $E_{\mu\nu}$,  for
a cosmological setting, we recover the generalized Friedmann
equation which contains one
integration constant and then it gives a closed form just as in the case of
Ref.~\cite{SMS}. This generalized Friedmann equation is the same as that 
obtained by the
previous authors\cite{GB_brane_cosmology}.
In this formulation, we need not to assume any functional form  for the
brane action. We can  add any
curvature terms in four dimensions, which may be induced by quantum effects of
matter fields. These brane-induced gravity models were investigated mainly in
the cosmological aspect
\cite{Dvali,Tanaka,Nojiri,Hawking}.

In Sec.~\ref{II}, we derive the covariant gravitational equations on  a
brane, applying it  to
cosmological model in Sec.~\ref{III}. We
obtain  the generalized Friedmann equation, which is given by a
cubic equation with respect to the Hubble parameter square $H^2$.
Conclusions and remarks follow in  Sec.~\ref{IV}.

\section{The effective gravitational equations}
\label{II}

We consider a 5-D bulk spacetime with
a single 4-D brane world, on which gravity is confined.
We assume the 5-D bulk spacetime  $({\cal M}, g_{AB})$, whose coordinates
are $X^A ~(A=0,1,2,3,5)$,
  is described by  the Einstein-Gauss-Bonnet action:
\begin{eqnarray}
S_{\rm bulk} =\int_{\cal M} d^5X \sqrt{- g} \left[
{1 \over 2 \kappa_5^2} \left({\cal R} +\alpha {\cal L}_{GB}\right)
+{\cal L}_{\rm m} \right],
\label{bulk_action}
\end{eqnarray}
where
\begin{eqnarray}
{\cal L}_{GB}={\cal R}^2-4{\cal R}_{AB}{\cal R}^{AB}
+{\cal R}_{ABCD}{\cal R}^{ABCD}.
\end{eqnarray}
$\kappa_5^2$ is the 5-D gravitational constant, ${\cal R}$,
${\cal R}_{AB}$,
${\cal R}_{ABCD}$
and ${\cal L}_{\rm m}$ are the 5-D scalar curvature, Ricci tensor,
Riemann  curvature and the matter
Lagrangian in the bulk, respectively.
$\alpha$ is a coupling constant.
The  4-D brane world $({\cal B}, h_{\mu\nu})$ is located
at a hypersurface
($\Sigma (X^A) = 0$) in the 5-D bulk
spacetime and
the induced 4-D metric $h_{\mu\nu}$ is defined by
\begin{eqnarray}
h_{AB} =g_{AB} - n_{A}n_{B} ,
\end{eqnarray}
where $n_A$ is the spacelike unit-vector field normal to the brane
hypersurface ${\cal B}$. The action is assumed to be given by the most
generic action:
\begin{eqnarray}
S_{\rm brane}=\int_{\cal B} d^4 x\sqrt{-h}
\left[ L_{\rm surface}
  + L_{\rm
brane}(h_{\alpha\beta},\psi)
\right]\,,
\label{brane_action}
\end{eqnarray}
where $x^\mu ~(\mu=0,1,2,3)$ are the induced 4-D  coordinates on
the  brane,
\begin{eqnarray}
L_{\rm surface}= {1\over\kappa_5^2}
\left[K+2\alpha(J-2G^{\rho\sigma}K_{\rho\sigma})\right]
\end{eqnarray}
is the surface term\cite{GH,ChaRea99,GB_brane_surface}, and $L_{\rm
brane}(h_{\alpha\beta},\psi)$ is the effective 4-D Lagrangian, which is 
given by a
generic functional of the brane metric $h_{\alpha\beta}$ and matter fields
$\psi$. $K_{\mu\nu} (= h_{~\mu}^{A} h_{~\nu}^{B} \nabla_A n_B) $,
  $K$, $J$ and $G^{\rho\sigma}$
  in the surface term are the extrinsic
curvature  of
${\cal B}$, its trace, its cubic
combination defined later, and the Einstein tensor of the induced metric
$h_{\mu\nu}$, respectively.

The total action
($S =  S_{\rm bulk}+ S_{\rm brane}$)
gives
our basic equations as
\begin{eqnarray}
{\cal G}_{AB} +\alpha {\cal H}_{AB}= \kappa_5^2 \,\, \left[
\,{\cal T}_{AB} +\tau_{AB}\, \delta(\Sigma)\,\right]
\, ,
\label{5dEinstein}
\end{eqnarray}
where
\begin{eqnarray}
{\cal G}_{AB}&=&{\cal R}_{AB}-{1\over 2}g_{AB}{\cal R},
\\
{\cal H}_{AB}&=&2\left[{\cal R}{\cal R}_{AB}-2{\cal R}_{AC}{\cal R}^C_{~B}
-2{\cal R}^{CD}{\cal R}_{ACBD}+
{\cal R}_A^{~CDE}{\cal R}_{BCDE}\right]-{1\over 2}g_{AB}{\cal L}_{GB},
\label{EGB:eq}
\end{eqnarray}
and
\begin{eqnarray}
{\cal T}_{AB} &\equiv &-2 {\delta {\cal L}_{\rm m} \over \delta
g^{AB}}  +g_{AB}{\cal L}_{\rm m}
\label{em_tensor_of_bulk}\,
\end{eqnarray}
is the energy-momentum tensor of
bulk matter fields,
while  $\tau_{\mu\nu}$ is the ``effective" energy-momentum tensor
localized  on the
brane which is defined by
\begin{eqnarray}
\tau_{\mu\nu}\equiv -2 {\delta L_{\rm
brane} \over \delta
h^{\mu\nu}}  +h_{\mu\nu}L_{\rm
brane}     \,.
\label{em_tensor_of_brane}
\end{eqnarray}
The $\delta(\Sigma)$ denotes the localization of brane contributions.
It is worth noting that  $\tau_{\mu\nu}$ may include curvature
contributions from induced gravity\cite{Dvali,MMT}. In that term, we can also
include ``non-local" contributions such as a trace
anomaly\cite{Nojiri,Hawking}, although those contributions are not
directly derived from  the effective
Lagrangian $L_{\rm brane}$.

The basic equations in the brane world are obtained by projecting
the variables onto the brane world  because we assume that the gravity
on the brane is confined.
We then project
the 5-D Riemann tensor  onto the brane spacetime as
\begin{eqnarray}
{\cal R}_{MNRS} ~h_{~A}^{M} \, h_{~B}^{N} \, h_{~C}^{R}\, h_{~D}^{S}
&= &
R_{ABCD}
      - K_{AC}K_{BD}
+K_{AD}K_{BC},
\label{Gauss}\\
{\cal R}_{MNRS} ~h_{~A}^{M} \, h_{~B}^{N} \, h_{~C}^{R}\, n^S
&= &2D_{[A}K_{B]C},
\label{Riemann_n} \\
{\cal R}_{MNRS} ~h_{~A}^{M}\,  h_{~C}^{R}\, n^{N}\, n^S
&= &
-\pounds_n K_{AC} +K_{AB}\, K^{B}_{~C}\, ,
\label{Riemann_nn}
\end{eqnarray}
where  $R_{ABCD}$ is the Riemann tensor of the induced metric $h_{MN}$,
  $D_M$ is
the covariant differentiation with respect to $h_{MN}$,
and $\pounds_n$ denotes the Lie derivative in the $n$-direction.
The first equation is called the Gauss equation.
Using this projection, the 5-D Riemann curvature and
its contractions (the Ricci tensor and scalar curvature) are described by
the 4-D variables on the brane
with the normal $n_M$
as
\begin{eqnarray}
{\cal R}_{MNRS}
&=&R_{MNRS}-K_{MR}K_{NS}+K_{MS}K_{NR}-n_M D_R K_{NS}+n_M
D_S K_{RN}+n_ND_RK_{SM}
-n_ND_SK_{RM}
\nonumber
\\
&&
-n_R D_MK_{NS}+n_RD_NK_{MS}+n_SD_MK_{NR}-n_SD_NK_{MR}
\nonumber
\\
&&
+n_Mn_R K_{NC}K^C_{~S}
-n_Mn_S K_{NC}K^C_{~R}
-n_Nn_R K_{MC}K^C_{~S}
+n_Nn_S K_{MC}K^C_{~R}
\nonumber
\\
&&-n_Mn_R {\pounds_n} K_{NS}
+n_Mn_S {\pounds_n} K_{NR}
+n_Nn_R {\pounds_n} K_{MS}
-n_Nn_S {\pounds_n} K_{MR},
\label{5DRiemann}
\\
{\cal R}_{MN}
&=&R_{MN}-KK_{MN}+2K_{MC}K^C_{~N}
+n_M \left(D_C K^C_{~N}- D_N K\right)
+n_N \left(D_C K^C_{~M}- D_M K\right)
\nonumber
\\
&&
+n_Mn_N K_{CD}K^{CD}
- {\pounds_n} K_{MN}
-n_Mn_N h^{CD}{\pounds_n} K_{CD},
\label{5DRicci}
\\
{\cal R}
&
=&R-K^2+3K_{CD}K^{CD}
-2 h^{CD}{\pounds_n} K_{CD}.
\label{5Dscalar}
\end{eqnarray}

As was shown by Deruelle and Madore\cite{GB_brane_linear},
the Einstein-Gauss-Bonnet
equation is quasi-linear, which means that apart from  non-singular terms 
given by the
4-dimensional variables,
it contains
only linear
terms of
   ${\pounds_n} K_{AB}$  but no quadratic terms appear.
In fact,
inserting  these relations into the basic equation (\ref{EGB:eq}),
   we find the effective equations on the brane as
\begin{eqnarray}
&&
M_{\mu\nu}-{1\over
2}Mh_{\mu\nu}+K_{\mu\rho}K^\rho_{~\nu}-h_{\mu\nu}K_{\alpha\beta}
K^{\alpha\beta}-{\pounds_n} K_{\mu\nu}+h_{\mu\nu}h^{\rho\sigma}
{\pounds_n} K_{\rho\sigma}
\nonumber \\
&&
~~+2\alpha\left(H_{\mu\nu}
-M{\pounds_n} K_{\mu\nu} +2M^{~\rho}_{\mu} {\pounds_n} K_{\rho\nu}
+2M^{~\rho}_{\nu} {\pounds_n} K_{\rho\mu} +W_{\mu\nu}^{~~~\rho\sigma}
{\pounds_n} K_{\rho\sigma}
\right)
=\kappa^2_5 \left[{\cal T}_{MN}h_{~\mu}^M
h_{~\nu}^N+\tau_{\mu\nu}\delta(\Sigma)
\right]
\,,
\label{eq:1}\\
&&
N_{\mu}+2\alpha \left(MN_\mu
-2M_\mu^{~\rho}N_\rho
+2M^{\rho\sigma}N_{\mu\rho\sigma}
-M_{\mu}^{~\nu\rho\sigma}N_{\rho\sigma\nu}
\right)
=\kappa^2_5 {\cal T}_{MN} n^N h_{~\mu}^M
\,,\label{eq:2}\\
&&M+\alpha\left(M^2-4M_{\alpha\beta}M^{\alpha\beta}
+M_{\alpha\beta\gamma\delta}M^{\alpha\beta\gamma\delta}\right)
=-2\kappa^2_5 {\cal T}_{MN} n^Mn^N
\,,\label{eq:3}
\end{eqnarray}
where
\begin{eqnarray}
M_{\alpha\beta\gamma\delta}
   &=&
R_{\alpha\beta\gamma\delta}-K_{\alpha\gamma}K_{\beta\delta}
+K_{\alpha\delta}K_{\beta\gamma},
\nonumber \\
M_{\alpha\beta}
&=&h^{\rho\sigma}M_{\alpha\rho\beta\sigma}=R_{\alpha\beta}
-KK_{\alpha\beta}+K_{\alpha\gamma}K^{\gamma}_{~\beta},
\nonumber \\
M &=&h^{\alpha\beta}M_{\alpha\beta}=R-K^2+K_{\alpha\beta}K^{\alpha\beta},
  \\
N_{\mu\nu\rho}
&=&
D_\mu K_{\nu\rho}-D_\nu K_{\mu\rho},
\nonumber \\
N_{\mu} &=&h^{\rho\sigma}N_{\rho\mu\sigma}=
D_\nu K_\mu^{~\nu}-D_\mu K,
\\
H_{\mu\nu} &=&
MM_{\mu\nu}-2(M_{\mu\rho}M^{\rho}_{~\nu}+M^{\rho\sigma}
M_{\mu\rho\nu\sigma} )+M_{\mu\rho\sigma\kappa}M_{\nu}^{~\rho\sigma\kappa}
+2K_{\alpha\beta}K^{\alpha\beta}M_{\mu\nu}+
MK_{\mu\rho}K^\rho_{~\nu}\nonumber \\
&&-2(K_{\mu\rho}K^\rho_{~\sigma}M^\sigma_{~\nu}
+K_{\nu\rho}K^\rho_{~\sigma}M^\sigma_{~\mu})
-2 K^{\rho\kappa}K_\kappa^{~\sigma}M_{\mu\rho\nu\sigma}
-2\left[N_\mu N_\nu-N^\rho (N_{\rho\nu\mu}+N_{\rho\mu\nu})\right]
\nonumber \\
&&
+N_{\rho\sigma\mu}N^{\rho\sigma}_{~~~\nu}
+2N_{\mu\rho\sigma}N^{~\rho\sigma}_{\nu}
-{1\over 4}h_{\mu\nu}\left[
M^2-4M_{\alpha\beta}M^{\alpha\beta}+M_{\alpha\beta\gamma\delta}
M^{\alpha\beta\gamma\delta}\right]
\nonumber \\
&&
+h_{\mu\nu}\left[-K_{\alpha\beta}K^{\alpha\beta}M
+2M_{\alpha\beta}K^{\alpha\gamma}K_{\gamma}^{~\beta}
+2N_\alpha N^\alpha -N_{\alpha\beta\gamma}N^{\alpha\beta\gamma}\right],
  \\
W_{\mu\nu}^{~~~\rho\sigma} &=&
Mh_{\mu\nu}h^{\rho\sigma}-2M_{\mu\nu}h^{\rho\sigma}-2h_{\mu\nu}
M^{\rho\sigma} +2M_{\mu\alpha\nu\beta}h^{\alpha\rho}h^{\beta\sigma}
\,.
\end{eqnarray}
Note that the linear
terms of
   ${\pounds_n} K_{CD}$ appear only in Eq. (\ref{eq:1}) but not in
Eqs. (\ref{eq:2}) and (\ref{eq:3}).
This is consistent with our ansatz that contribution from a brane
is given by Eq. (\ref{em_tensor_of_brane}) because its
contraction by
$n_A$ vanishes.

The singular behavior in 5-D bulk spacetime appears
in the 5-D gravitational equations (\ref{5dEinstein})
as the $\delta$-function.
Then,  ${\pounds_n} K_{CD}$ has the $\delta$-functional
singularity to balance to the energy-momentum tensor of the brane
world.
Integrating Eq.~(\ref{eq:1}) in the $n$-direction, we obtain
the generalized Israel's junction condition\cite{israel,GB_brane_junction};
\begin{eqnarray}
[K_{\mu\nu}]_\pm-h_{\mu\nu}[K]_\pm +2\alpha\Bigr(
3[J_{\mu\nu}]_\pm -h_{\mu\nu}[J]_\pm  -2
P_{\mu\rho\nu\sigma}[K^{\rho\sigma}]_\pm
\Bigr) =-\kappa_5^2 \tau_{\mu\nu}\,,
\end{eqnarray}
where
\begin{eqnarray}
J_{\mu\nu} &=&{1\over 3}\left(
2KK_{\mu\rho}K^{\rho}_{~\nu}
+K_{\rho\sigma}K^{\rho\sigma}K_{\mu\nu}
-2K_{\mu\rho}K^{\rho\sigma}K_{\sigma \nu}
-K^2 K_{\mu\nu}
\right)
\\
P_{\mu\nu\rho\sigma}&=&
R_{\mu\nu\rho\sigma}+2h_{\mu[\sigma}R_{\rho]\nu}
+2h_{\nu[\rho}R_{\sigma]\mu} +Rh_{\mu[\rho}h_{\sigma]\nu}
\,.
\end{eqnarray}
We have introduced
\begin{equation}
[X]_\pm\equiv X^+-X^- ,
\end{equation}
  where $X^\pm$ are $X$'s
evaluated either on the
$+$ or $-$ side of the brane and
$P_{\mu\nu\rho\sigma}$ is the divergence free part of the Riemann tensor,
i.e.
\begin{equation}
D_\mu P^{\mu}_{~\nu\rho\sigma}=0.
\end{equation}

Because of the $Z_2$-symmetry,
we have
\begin{equation}
K_{\mu\nu}^+=-K^-_{\mu\nu}\, ,
\label{eq:zsym}
\end{equation}
then the
extrinsic curvature of the brane is uniquely determined
by the junction condition as
\begin{eqnarray}
B_{\mu\nu}=-{\kappa_5^2\over 2} \tau_{\mu\nu}\,,
\label{junction}
\end{eqnarray}
where
\begin{eqnarray}
B_{\mu\nu}\equiv K_{\mu\nu}-K h_{\mu\nu}+2\alpha\left(
3J_{\mu\nu} -Jh_{\mu\nu}-2 P_{\mu\rho\nu\sigma}K^{\rho\sigma} \right).
\end{eqnarray}
In what follows, we  omit the indices $\pm$ below for brevity.

The above quasi-linearity guarantees the ansatz of an infinitely
thin brane. The obtained equations for induced metric is described by
geometrical quantities and does not depend on microphysics of the brane.
This situation will be changed when  we discuss other curvature-squared
terms. On the other hand, if we include the Lovelock Lagrangian which is
higher than Gauss-Bonnet one but does not contain higher-derivatives,
we can assume that a brane is infinitely thin because it is also
quasi-linear and then extend the present
approach.

In order to find the effective equations on the brane, we
have to replace the  terms of ${\pounds_n} K_{\mu\nu}$ in Eq. (\ref{eq:1})
with  the 4-D variables on the brane.
The singular part in Eq. (\ref{eq:1}) has been evaluated by the  junction
condition.
Hence
  we have to  evaluate  ${\pounds_n} K_{\mu\nu}$  either on the
$+$ or $-$ side of the brane, which is nonsingular.
 From Eqs. (\ref{Riemann_nn}), (\ref{5DRicci}) and (\ref{5Dscalar}) with
the decomposition of the Riemann tensor as
\begin{eqnarray}
{\cal R}_{ABCD}
=\frac{2}{3}\left(g_{A [C}{\cal R}_{D]B}
-g_{B [ C}{\cal R}_{D] A}\right)
-\frac{1}{6} g_{A [C} g_{D ]B}{\cal R}
+{\cal C}_{ABCD}\, ,
\label{Riemann}
\end{eqnarray}
where ${\cal C}_{MRNS}$ is the 5-D Weyl curvature,
we find
\begin{eqnarray}
{\pounds_n} K_{\mu\nu}-{1\over 4} h_{\mu\nu} h^{\alpha\beta}{\pounds_n}
K_{\alpha\beta}=-{3\over 2} E_{\mu\nu}-{1\over 2}\left(
M_{\mu\nu}-{1\over 4}h_{\mu\nu}M\right)
+K_{\mu\rho}K^{\rho}_{~\nu}
-{1\over 4} h_{\mu\nu}K_{\rho\sigma}K^{\rho\sigma}\,,
\label{LKMN1}
\end{eqnarray}
where
\begin{equation}
E_{\mu\nu} \equiv {\cal C}_{MRNS}~n^M n^N
h_{~\mu}^{R}~ h_{~\nu}^{S} \,.
\label{Edef}
\end{equation}
However, because  Eq. (\ref{LKMN1}) is a trace free equation,
we cannot fix ${\pounds_n} K_{\mu\nu}$ by Eq. (\ref{LKMN1}).
We have to find $h^{\alpha\beta}{\pounds_n}
K_{\alpha\beta}$ from other independent equation.
We shall take a trace of our basic equation (\ref{5dEinstein}), finding
\begin{eqnarray}
3{\cal R} +\alpha{\cal L}_{GB}=-2\kappa_5^2{\cal T}.
\label{trace_5dEinstein}
\end{eqnarray}
Inserting Eqs. (\ref{Gauss})-(\ref{Riemann_nn}) with
Eq. (\ref{LKMN1}) into Eq. (\ref{trace_5dEinstein}),
we find
\begin{eqnarray}
h^{\alpha\beta}{\pounds_n}
K_{\alpha\beta}={M\over
2}+K_{\alpha\beta}K^{\alpha\beta}+{\kappa_5^2\over 3+\alpha M}
{\cal T} +{\alpha\over 2(3+\alpha M)} I,
\label{traceLKMN}
\end{eqnarray}
where
\begin{eqnarray}
I=
M^2-8M_{\alpha\beta}M^{\alpha\beta}+M_{\alpha\beta\gamma\delta}
M^{\alpha\beta\gamma\delta}
-8N_\rho N^\rho +4 N_{\rho\sigma\kappa}N^{\rho\sigma\kappa}
-12M_{\rho\sigma}E^{\rho\sigma}\,.
\end{eqnarray}
 From Eq.  (\ref{LKMN1})  with Eq. (\ref{traceLKMN}), we then find
\begin{eqnarray}
{\pounds_n}
K_{\mu\nu}=
-{3\over 2} E_{\mu\nu}
-{1\over 2}\left(M_{\mu\nu}-{1\over 2}Mh_{\mu\nu}\right)
+K_{\mu\rho}K^{\rho}_{\nu}
+{\kappa_5^2\over 4(3+\alpha M)}{\cal T} h_{\mu\nu}
  +{\alpha\over 8(3+\alpha M)}I h_{\mu\nu}.
\label{LKMN2}
\end{eqnarray}
Inserting Eq. (\ref{LKMN2}) into Eq. (\ref{eq:1}), we  obtain
the effective gravitational equations on the brane
as
\begin{eqnarray}
&&
{3\over 2}\left(M_{\mu\nu}+E_{\mu\nu}\right)-{1\over 4}M
h_{\mu\nu}
+\alpha\left[H^{(1)}_{\mu\nu}+H^{(2)}_{\mu\nu}
+H^{(3)}_{\mu\nu}\right]-{\alpha^2I\over
2(3+\alpha M)}\left( M_{\mu\nu}-{1\over 4}Mh_{\mu\nu}\right)\nonumber
\\ &&~~~~~=\kappa^2_5
\left[{\cal T}_{MN}h_{~\mu}^M h_{~\nu}^N-{1\over 4}h_{\mu\nu}{\cal
T}
+{\alpha\over 3+\alpha M}\left(M_{\mu\nu}-{1\over 4}M
h_{\mu\nu}\right){\cal T}\right]
\,,
\label{eq:brane1}
\end{eqnarray}
where
\begin{eqnarray}
H^{(1)}_{\mu\nu}
&=&2M_{\mu\alpha\beta\gamma}M_{\nu}^{~\alpha\beta\gamma}-6M^{\rho\sigma}
M_{\mu\rho\nu\sigma}+4MM_{\mu\nu}-8M_{\mu\rho}M_{\nu}^{~\rho}
-{1\over
8}h_{\mu\nu}\left(7M^2-24M_{\alpha\beta}M^{\alpha\beta}
+3M_{\alpha\beta\gamma\delta}M^{\alpha\beta\gamma\delta}\right),
\nonumber \\
H^{(2)}_{\mu\nu}
&=&-6\left(M_{\mu\rho}E^\rho_{~\nu}+M_{\nu\rho}E^\rho_{~\mu}+
M_{\mu\rho\nu\sigma}E^{\rho\sigma}
\right)+{9\over
2}h_{\mu\nu}M_{\rho\sigma}E^{\rho\sigma} +3ME_{\mu\nu},
\nonumber \\
H^{(3)}_{\mu\nu}
&=&-4N_\mu N_\nu +4N^\rho\left(N_{\rho\mu\nu}+N_{\rho\nu\mu}\right)
+2N_{\rho\sigma\mu}N^{\rho\sigma}_{~~~\nu}
+4N_{\mu\rho\sigma}N_{\nu}^{~\rho\sigma}+
3h_{\mu\nu}\left(N_\alpha N^{\alpha}-{1\over
2}N_{\alpha\beta\gamma}N^{\alpha\beta\gamma}\right)
\,.
\end{eqnarray}

Eq. (\ref{eq:3}) is automatically satisfied when we take a trace of Eq.
(\ref{eq:brane1}),
which means that it is not independent.
We have then two basic
equations (\ref{eq:brane1}) and (\ref{eq:2}). Eq. (\ref{eq:2}) is
rewritten as
\begin{eqnarray}
D_\nu \Bigl[K_\mu^{~\nu}-K\delta_\mu^{~\nu}
+2\alpha (3J_\mu^{~\nu}-J\delta_\mu^{~\nu}
-2P_{\mu}^{~\rho\nu\sigma}K_{\rho\sigma}
)\Bigr] =\kappa_5^2{\cal
T}_{MN}h^M_\mu n^N
\,,
\end{eqnarray}
which gives
the constraint on the brane matter fields through the junction condition
(\ref{junction}), i.e.
\begin{eqnarray}
D_\nu \tau_\mu^{~\nu}=-2{\cal
T}_{MN}h^M_\mu n^N\,.
\end{eqnarray}
If there is no energy-momentum transfer from the bulk, we find the
energy-momentum conservation of  brane matter fields as
\begin{eqnarray}
D_\nu \tau_\mu^{~\nu}=0\,.
\label{EM_cons}
\end{eqnarray}

In Eq. (\ref{eq:brane1}),
we have so far three unknown variables;
${\cal T}_{AB}$, $E_{\mu\nu}$, and $K_{\mu\nu} $.
The first two variables are described by bulk information,
whereas  the extrinsic curvature $K_{\mu\nu}$
is related to the
  brane ``energy-momentum" tensor $\tau_{\mu\nu}$ as
Eq. (\ref{junction}).
Hence
Eqs. (\ref{junction}) and (\ref{eq:brane1}) with the energy momentum
conservation (\ref{EM_cons})
give the effective gravity theory on the
brane.

It may be better to rewrite Eq. (\ref{eq:brane1}) to the Einstein-type 
equations
with ``correction" terms. From Eqs. (\ref{eq:3}) and
(\ref{eq:brane1}), we find
\begin{eqnarray}
&&
G_{\mu\nu}+E_{\mu\nu}-KK_{\mu\nu}+K_{\mu\rho}K^\rho_{~\nu}+{1\over
2}
\left(
K^2-K_{\alpha\beta}K^{\alpha\beta}
\right)
h_{\mu\nu}+\alpha\left(\hat{H}^{(1)}_{\mu\nu}+\hat{H}^{(2)}_{\mu\nu}
+\hat{H}^{(3)}_{\mu\nu}\right)
\nonumber
\\ &&~~~=
{2\kappa^2_5\over 3}\left\{
\left[
{\cal T}_{MN}h_{~\mu}^M h_{~\nu}^N+
\left(
{\cal
T}_{MN}n^{M}n^{N}
-{1\over 4}{\cal
T}^M_{~M}
\right)
h_{M\mu\nu}
\right]
+ {\alpha\over 3+\alpha M}\left(M_{\mu\nu}-{1\over
4}Mh_{\mu\nu}\right){\cal T}_{MN}h^{MN}\right\}
\,,
\label{eq:brane_Etype}
\end{eqnarray}
where
\begin{eqnarray}
\hat{H}^{(1)}_{\mu\nu}&=&
{4\over
3}\left(M_{\mu\alpha\beta\gamma}M_{\nu}^{~\alpha\beta\gamma}-3M^{\rho\sigma}
M_{\mu\rho\nu\sigma}+2MM_{\mu\nu}-4M_{\mu\rho}M_{\nu}^{~\rho}\right)
  -{1\over
12}h_{\mu\nu}\left(11M^2-40M_{\alpha\beta}M^{\alpha\beta}
+7M_{\alpha\beta\gamma\delta}M^{\alpha\beta\gamma\delta}\right)
\nonumber \\
&&-{\alpha\over 3(3+\alpha M)}\left(M_{\mu\nu}-{1\over 4}Mh_{\mu\nu}\right)
\left(M^2-8M_{\alpha\beta}M^{\alpha\beta}
+M_{\alpha\beta\gamma\delta}M^{\alpha\beta\gamma\delta}\right),
\nonumber \\
\hat{H}^{(2)}_{\mu\nu}&=&-4\left(M_{\mu\rho}E^\rho_{~\nu}
+M_{\nu\rho}E^\rho_{~\mu}+
M_{\mu\rho\nu\sigma}E^{\rho\sigma}
\right)+3h_{\mu\nu}M_{\rho\sigma}E^{\rho\sigma} +2ME_{\mu\nu}
+{4\alpha\over 3+\alpha M}\left(M_{\mu\nu}-{1\over
4}Mh_{\mu\nu}\right)M_{\rho\sigma}E^{\rho\sigma},
\nonumber \\
\hat{H}^{(3)}_{\mu\nu}&=&{8\over
3}\left[-N_\mu N_\nu +N^\rho\left(N_{\rho\mu\nu}+N_{\rho\nu\mu}\right) +
{1\over
2}N_{\rho\sigma\mu}N^{\rho\sigma}_{~~~\nu}
+N_{\mu\rho\sigma}N_{\nu}^{~\rho\sigma}\right]\nonumber \\
&&+
\left[2h_{\mu\nu}+{8\alpha\over 3(3+\alpha M)}\left(M_{\mu\nu}-{1\over
4}Mh_{\mu\nu}\right)\right]\left(N_\alpha N^{\alpha}-{1\over
2}N_{\alpha\beta\gamma}N^{\alpha\beta\gamma}\right).
\end{eqnarray}
As for the junction condition,
we find
\begin{eqnarray}
&&K_{\mu\nu}+{2\alpha\over 3}\Bigl[9J_{\mu\nu}
-2Jh_{\mu\nu}-2\left(3P_{\mu\rho\nu\sigma}+h_{\mu\nu} G_{\rho\sigma}
\right)K^{\rho\sigma}
\Bigr]
=-{\kappa_5^2\over 2}\left(\tau_{\mu\nu}-{1\over 3}\tau h_{\mu\nu}
\right)
\label{junction_Etype}
\,.
\end{eqnarray}

If $\alpha=0$, we find two equations:
\begin{eqnarray}
&&
G_{\mu\nu}+E_{\mu\nu}-KK_{\mu\nu}+K_{\mu\rho}K^\rho_{~\nu}+{1\over
2}
\left(
K^2-K_{\alpha\beta}K^{\alpha\beta}
\right)
h_{\mu\nu}
\nonumber \\
&&~~~~~~
={2\kappa^2_5\over 3}
\left[
{\cal T}_{MN}h_{~\mu}^M h_{~\nu}^N+
\left(
{\cal
T}_{MN}n^{M}n^{N}
-{1\over 4}{\cal
T}^M_{~M}
\right)
h_{M\mu\nu}
\right]
\,,
\label{eq:brane_alpha0}
\\
&&K_{\mu\nu}=-{\kappa_5^2\over 2}\left(\tau_{\mu\nu}-{1\over 3}\tau
h_{\mu\nu}
\right)
\,,
\end{eqnarray}
which are exactly the same as those found in
Ref.~\cite{SMS}, which gives  the Einstein gravitational theory in the 4-D
brane world.
However, if the Gauss-Bonnet term appears,
gravitational interaction  on the brane will be modified
in the effective theory.
 
The gravity on the brane is described by Eq.  (\ref{eq:brane1}) with Eq.
(\ref{junction}), or equivalently by Eq. (\ref{eq:brane_Etype}) with Eq.
(\ref{junction_Etype}). Just as the case of the RS II model, this system is 
not closed
because of  appearance of the terms with $E_{\mu\nu}$, which is
some component of the 5-D Weyl curvature.
Although we have to solve a bulk spacetime as well as a brane world,
we know that any contribution from a bulk spacetime to a brane world
is described only by the tidal force $E_{\mu\nu}$.

Although the above form  (\ref{eq:brane1}) or (\ref{eq:brane_Etype}) is 
good enough to
describe our basic equations, it is sometimes convenient to divide Eq. 
(\ref{eq:brane1})
into two parts; its trace and the trace free equation. Introducing trace 
free variables
as
\begin{eqnarray}
\tilde{M}_{\mu\nu}&=& M_{\mu\nu}-{1\over 4}Mh_{\mu\nu},\nonumber \\
L_{\mu\nu\rho\sigma}&=& M_{\mu\nu\rho\sigma}+h_{\mu
[\sigma}\tilde{M}_{\rho]\nu} +h_{\nu [\rho}\tilde{M}_{\sigma]\mu}
  -{1\over
6}Mh_{\mu [\rho}h_{\sigma]\nu}\,,
\end{eqnarray}
we find
\begin{eqnarray}
&&M+\alpha\left({1\over 6}M^2
-2\tilde{M}_{\alpha\beta}\tilde{M}^{\alpha\beta}
+L_{\alpha\beta\gamma\delta}L^{\alpha\beta\gamma\delta}\right)
=-2\kappa^2_5 {\cal T}_{MN} n^Mn^N,
\label{eq:brane_trace}\\
&&
{3\over 2}\left(\tilde{M}_{\mu\nu}+E_{\mu\nu}\right)
+\alpha\left[\bar{H}^{(1)}_{\mu\nu}+\bar{H}^{(2)}_{\mu\nu}
+\bar{H}^{(3)}_{\mu\nu}\right]
\nonumber
\\ &&~~~~~=\kappa^2_5
\left[{\cal T}_{MN}h_{~\mu}^M h_{~\nu}^N-{1\over 4}h_{\mu\nu}
{\cal
T}_{MN}h^{MN}
+{\alpha\over 3+\alpha M}\tilde{M}_{\mu\nu}{\cal
T}_{MN}h^{MN}\right]
\,,
\label{eq:brane_tracefree}
\end{eqnarray}
where
\begin{eqnarray}
\bar{H}^{(1)}_{\mu\nu}&=&
2\left(L_{\mu\alpha\beta\gamma}L_{\nu}^{~~\alpha\beta\gamma}
-\tilde{M}^{\alpha\beta}L_{\mu\alpha\nu\beta}
-\tilde{M}_{\mu}^{~\alpha}\tilde{M}_{\alpha\nu}\right)-{3-\alpha M\over
6(3+\alpha M)} M\tilde{M}_{\mu\nu}
+{2\alpha\over 3+\alpha
M}\tilde{M}_{\alpha\beta}\tilde{M}^{\alpha\beta}\tilde{M}_{\mu\nu}
\nonumber \\
&& -{1\over
2}h_{\mu\nu}\left(L_{\alpha\beta\gamma\delta}L^{\alpha\beta\gamma\delta}
-\tilde{M}_{\alpha\beta}\tilde{M}^{\alpha\beta}
\right),
\nonumber
\\
\bar{H}^{(2)}_{\mu\nu}&=&
-3\left(\tilde{M}_{\mu\rho}E^\rho_{~\nu}
+\tilde{M}_{\nu\rho}E^\rho_{~\mu}+
2L_{\mu\rho\nu\sigma}E^{\rho\sigma}
\right)+{3\over
2}h_{\mu\nu}\tilde{M}_{\rho\sigma}E^{\rho\sigma} +{1\over 2}ME_{\mu\nu}
+{6\alpha\over 3+\alpha
M}\tilde{M}_{\rho\sigma}E^{\rho\sigma}\tilde{M}_{\mu\nu},\nonumber
\\
\bar{H}^{(3)}_{\mu\nu}&=& -4N_\mu N_\nu
+4N^\rho\left(N_{\rho\mu\nu}+N_{\rho\nu\mu}\right)
+2N_{\rho\sigma\mu}N^{\rho\sigma}_{~~~\nu}
+4N_{\mu\rho\sigma}N_{\nu}^{~\rho\sigma}+
3h_{\mu\nu}\left(N_\alpha N^{\alpha}-{1\over
2}N_{\alpha\beta\gamma}N^{\alpha\beta\gamma}\right)
\nonumber \\
&&
+{4\alpha\over 3+\alpha M}\left(N_\alpha N^{\alpha}-{1\over
2}N_{\alpha\beta\gamma}N^{\alpha\beta\gamma}\right)\tilde{M}_{\mu\nu}
\,.
\end{eqnarray}

The junction condition  (\ref{junction}) is also
decomposed
into two parts:
\begin{eqnarray}
B&\equiv&B^\mu_{~\mu}\nonumber \\
&=&-3K+\alpha\left(
4\tilde{M}_{\rho\sigma}\tilde{K}^{\rho\sigma}-KM-{1\over 2}K^3+2K
\tilde{K}_{\rho\sigma}\tilde{K}^{\rho\sigma}
-{8\over
3}\tilde{K}^{\rho}_{~\sigma}\tilde{K}^{\sigma}_{~\kappa}
\tilde{K}^{\kappa}_{~\rho}\right)=-{\kappa_5^2\over 2}\tau,
\label{junction_trace}\\
\tilde{B}_{\mu\nu}&\equiv &B_{\mu\nu}-{1\over 4}B
h_{\mu\nu}\nonumber \\
&=&
\tilde{K}_{\mu\nu}-12\alpha\tilde{J}_{\mu\nu}
-\alpha\left[
4
L_{\mu\rho\nu\sigma}\tilde{K}^{\rho\sigma}
+2\tilde{K}_{\mu}^{~\rho}\tilde{M}_{\rho\nu}
+2\tilde{K}_{\nu}^{~\rho}\tilde{M}_{\rho\mu}
-h_{\mu\nu}\tilde{K}_{\rho\sigma}\tilde{M}^{\rho\sigma}
-K\tilde{M}_{\mu\nu}-{1\over 3}M\tilde{K}_{\mu\nu}
\right]
\nonumber \\
&=&
-{\kappa_5^2\over 2}\left(\tau_{\mu\nu}-{1\over 4}\tau h_{\mu\nu}
\right)\,,
\label{junction_tracefree}
\end{eqnarray}
where
\begin{eqnarray}
\tilde{K}_{\mu\nu}&=&K_{\mu\nu}-{1\over 4}Kh_{\mu\nu},
\nonumber \\
\tilde{J}_{\mu\nu}&=&J_{\mu\nu}-{1\over 4}Jh_{\mu\nu}
\nonumber \\
&=&{1\over
3}\left[-2\tilde{K}_{\mu\rho}\tilde{K}^{\rho\sigma}\tilde{K}_{\sigma\nu}
+{1\over 2}K\tilde{K}_{\mu\rho}\tilde{K}^{\rho}_{~\nu}
+\tilde{K}_{\mu\nu}\left(\tilde{K}_{\rho\sigma}\tilde{K}^{\rho\sigma}
-{1\over 8}K^2\right)\right.
\nonumber
\\
&&
~~~~~
\left.
+{1\over 2}h_{\mu\nu}\left(
\tilde{K}^{\alpha}_{~\beta}\tilde{K}^{\beta}_{~\gamma}
\tilde{K}^{\gamma}_{~\alpha}-{1\over
4}K\tilde{K}_{\rho\sigma}\tilde{K}^{\rho\sigma}\right)\right]
\,.
\end{eqnarray}

In this decomposition, our basic equations are (\ref{eq:brane_trace}) and
(\ref{eq:brane_tracefree}) with the junction conditions 
(\ref{junction_trace}) and
(\ref{junction_tracefree}).  This form may be better to describe some
symmetric spacetime such as the Friedmann-Robertson-Walker (FRW)  universe, 
which we
shall study next.

\section{Friedmann equation}
\label{III}
We apply the present reduction to the FRW
cosmology. We assume spacetime as
\begin{eqnarray}
ds^2 =-dt^2 +a^2(t)~ \gamma_{ij}~ dx^i dx^j\,,
\end{eqnarray}
where $\gamma_{ij}$  denotes the metric of maximally symmetric 
3-dimensional space.
This gives
\begin{eqnarray}
&&R^0_{~0}=3(X+Y), ~~~~~~~~~R^i_{~j} =(Y+3X)\delta^i_{~j},
\nonumber \\
&&R^{0i}_{~~0j}=(X+Y) \delta^i_{~j},~~~~R^{ij}_{~~kl}=X
\left(
\delta^i_{~k}\delta^j_{~l}-\delta^i_{~l}\delta^j_{~k}
\right),
\nonumber \\
&&P^{0i}_{~~0j}=X \delta^i_{~j},~~~~~~~~~~~~~P^{ij}_{~~kl}=(X+Y)
\left(
\delta^i_{~k}\delta^j_{~l}-\delta^i_{~l}\delta^j_{~k}
\right)\,,
\end{eqnarray}
where
\begin{eqnarray}
X\equiv H^2+{k\over a^2},
~~~~~
Y\equiv \dot{H}-{k\over a^2}\,.
\end{eqnarray}

We assume that only a cosmological constant exists in the bulk, i.e.,
\begin{eqnarray}
\kappa_5^2 {\cal T}_{MN}=-\Lambda g_{MN}.
\end{eqnarray}
 From the symmetry of FRW spacetime,
we can set
\begin{eqnarray}
K^\mu_{~\nu}&=& \left(K^0_{~0},{\cal K} \delta^i_{~j}\right)\, ,
\\
E^\mu_{~\nu}
&=&E^0_{~0}\left(1, -{1\over 3}\delta^i_{~j}\right)\,.
\end{eqnarray}
We note that $E^\mu_{~\nu}$ is trace free.
We then write down the basic equations (\ref{eq:brane_trace}) and
(\ref{eq:brane_tracefree}) as
\begin{eqnarray}
&&2\bar{X}+\bar{Y}+4\alpha\bar{X}(\bar{X}+\bar{Y})
={\Lambda\over
3},
\label{eq00_F}
\\
&&3\bar{Y}+2E^0_{~0}+\alpha\left(H^{(1)}+H^{(2)}\right)=-{8\alpha\Lambda
\over 3+\alpha
M} \bar{Y}\,,
\label{eq_tracefree_F}
\end{eqnarray}
where
\begin{eqnarray}
&&
\bar{X}= X-{\cal K}^2,
\,\nonumber
\\ &&
\bar{Y}=Y-K^0_{~0}{\cal K}+{\cal K}^2,
\\
&& M=6(2\bar{X}+\bar{Y}),
\,\nonumber
\\&&
H^{(1)}\equiv -2\bar{Y}\left[2\bar{X}+3\bar{Y}-{\alpha\over
3(3+\alpha M)}\left(M^2+18\bar{Y}^2\right)\right],
\,\nonumber
\\ &&
H^{(2)}\equiv 4\left(2\bar{X}-\bar{Y}+{6\alpha\over 3+\alpha
M}\bar{Y}^2\right)E^0_{~0}
\,.
\end{eqnarray}
Here we have used the fact that
$L_{\mu\nu\rho\sigma}$ vanishes and
\begin{eqnarray}
\tilde{M}^\mu_{~\nu}={3\over 2}\bar{Y} \left(1, -{1\over
3}\delta^i_{~j}\right).
\end{eqnarray}
We have also found that $N_\rho$ and $N_{\alpha\beta\gamma}$ vanish
as will be shown below.
The  non-trivial components of $N_\rho$ and
$N_{\alpha\beta\gamma}$ are calculated as
\begin{eqnarray}
N_0=-3{\cal N},
~~~~~N_{0i}^{~~j}=-N_{i0}^{~~j}={\cal N}\delta_i^{~j}\,,
\end{eqnarray}
where
${\cal N}\equiv \dot{\cal K}-H(K^0_{~0}-{\cal K})$,
which gives
the l.h.s. of Eq. (\ref{eq:2}) as
\begin{eqnarray}
-3{\cal N}(1+4\alpha\bar{X})\,.
\end{eqnarray}
Since the r.h.s. of Eq. (\ref{eq:2}) vanishes when only
a cosmological
constant appears in a bulk spacetime, we obtain
\begin{eqnarray}
{\cal N}=\dot{\cal K}-H(K^0_{~0}-{\cal K})=0\,,
\label{em_cons0}
\end{eqnarray}
resulting that $N_\rho=N_{\alpha\beta\gamma}=0$.
 
As for the junction condition (\ref{junction}),
we first  calculate $J_{\mu\nu}$ as
\begin{eqnarray}
&&\tilde{J}^\mu_{~\nu}=-{1\over 2}{\cal K}^2 (K^0_{~0}-{\cal K})
\left(1,-{1\over 3}\delta^i_{~j}\right),\,\nonumber \\
&&J=-2{\cal K}^2 \left(3K^0_{~0}+{\cal K}
\right)\,,
\end{eqnarray}
and then obtain
\begin{eqnarray}
&&\tilde{B}^\mu_{~\nu}={3\over 4}\Bigl[(K^0_{~0}-{\cal
K})(1+4\alpha\bar{X}+8\alpha{\cal K}^2)+8\alpha{\cal
K}\bar{Y}\Bigr]\left(1,-{1\over 3}\delta^i_{~j}\right),
\,\nonumber \\
&&B=-3(K^0_{~0}+3{\cal K})-4\alpha
  \Bigl[
2{\cal K}^2({\cal K}+3K^0_{~0})+3(K^0_{~0}+3{\cal K})\bar{X}
+6{\cal K}\bar{Y}
\Bigr]\,.
\end{eqnarray}
The junction condition  (\ref{junction}) gives two independent relations
\begin{eqnarray}
&&{\cal K}(1+4\alpha \bar{X})+{8\over 3}\alpha {\cal
K}^3={\kappa_5^2 \over 6}\tau^0_{~0},\,
\label{junc_cond1}
\\ &&
(K^0_{~0}-{\cal K})\bigl[1+4\alpha(\bar{X}+2{\cal K}^2)\bigr]+8\alpha {\cal
K}\bar{Y}={\kappa_5^2\over 2}(\tau^1_{~1}-\tau^0_{~0})\,.
\label{junc_cond2}
\end{eqnarray}
We also find
\begin{eqnarray}
\dot{\bar{X}}=2H\bar{Y}\,
\label{dotbarX}
\end{eqnarray}
from  Eq. (\ref{em_cons0}) with $\dot{X}=2HY$.
We then recover the energy-momentum conservation law
\begin{eqnarray}
\dot{\tau}^0_{~0}+3H(\tau^0_{~0}-\tau^1_{~1})=0\,.
\end{eqnarray}
from Eqs. (\ref{junc_cond1}) and (\ref{junc_cond2}).

Using Eq. (\ref{dotbarX}) with Eq. (\ref{eq00_F}),
we obtain
\begin{eqnarray}
\dot{\bar{X}}=2H \left(
1+4\alpha \bar{X}\right)^{-1}\left[
{\Lambda\over
3}-2\bar{X}\left(1+2\alpha\bar{X}\right)\right]
\,,
\end{eqnarray}
which is easily integrated as
\begin{eqnarray}
\bar{X}(1+2\alpha\bar{X})={\Lambda\over 6} +{C\over a^4}\,.
\end{eqnarray}
where $C$ is an integration constant.

If $\alpha=0$, we have
\begin{eqnarray}
\bar{X}=\bar{X}_0(a)\equiv {\Lambda\over 6} +{C\over a^4}\,.
\label{X_0}
\end{eqnarray}
With Eq.  (\ref{junc_cond1}) we find
\begin{eqnarray}
X\equiv\bar{X}_0(a)+{\cal K}^2= {\kappa_5^4 \over 36}(\tau^0_{~0})^2
+{\Lambda\over
6} +{C\over a^4}\,.
\end{eqnarray}
Setting $\tau^0_{~0}=-(\lambda+\rho)$, where
$\lambda$  is a positive tension of the brane and $\rho$ is the energy density,
  we recover the
well-known equation in brane cosmology, i.e.
\begin{eqnarray}
H^2+{k\over a^2}={\Lambda_4\over 3}+{8\pi G\over 3}\rho +{\kappa_5^4\over
36}\rho^2+ {E^0_{~0}\over 3}
\,,
\end{eqnarray}
where
\begin{eqnarray}
\Lambda_4={1\over 2} \left(\Lambda+{\kappa_5^4\lambda^2\over
6}\right)\,,
~~~~~~~~
8\pi G={\kappa_5^4\over 6}\lambda
\,.
\end{eqnarray}
We have used  $E^0_{~0}(=-3\bar{Y}/2)=3C/a^4$, which is obtained from Eqs.
(\ref{eq00_F}), (\ref{eq_tracefree_F}), and (\ref{X_0}).

When $\alpha\neq 0$,
solving the above quadratic equation, we obtain
\begin{eqnarray}
\bar{X}=\bar{X}_\pm (a)\equiv {1\over 4\alpha}\left[
-1\pm\sqrt{1+8\alpha\bar{X}_0(a)}\right]\,.
\label{X_pm}
\end{eqnarray}
Inserting
$
{\cal K}^2 =X-\bar{X}_\pm (a)\,
$
into the square of Eq. (\ref{junc_cond1}), we find
\begin{eqnarray}
\left[X-\bar{X}_\pm (a)\right]\left[
1+{8\over 3}\alpha X +{4\over 3}\alpha\bar{X}_\pm (a)\right]^2
={\kappa_5^4\over 36}\left(\tau^0_{~0}\right)^2
\label{generalized_Friedmann}\,.
\end{eqnarray}
This is
generalization of Friedmann equation.
If the brane contains only matter field including a tension, i.e.
$\tau^0_{~0}=-(\lambda +\rho)$,
  it is a cubic
equation with respect to
$X=H^2+k/a^2$\cite{GB_brane_cosmology,GB_brane_junction}. When we have the
Einstein-Hilbert action on the brane such as an induced
gravity\cite{MMT,Dvali}, the generalized Friedmann equation becomes 
complicated, but it is still a cubic equation\cite{GB_brane_cosmology}.
In the case with a trace anomaly, $\tau^0_{~0}$ contains not only
$\lambda, \rho$, and $X$, but also $Y$ and $\dot{Y}$.
As a result, we have a very complicated equation.

The other independent equation (\ref{eq_tracefree_F}) just gives the value of
$E^0_{~0}$, i.e.
\begin{eqnarray}
E^0_{~0}=-{3\over 2}\bar{Y}_\pm(a)
\left[1+{8\alpha\bar{Y}_\pm(a) \over
3(1+4\alpha\bar{X}_\pm(a))}\right]
\,,
\label{E_OO}
\end{eqnarray}
where
\begin{eqnarray}
\bar{Y}_\pm (a)&=&\left(1+4\alpha \bar{X}_\pm(a)\right)^{-1}
\left[{\Lambda\over 3}-2\bar{X}_\pm(a)
\left(1+2\alpha\bar{X}_\pm(a)\right)\right]
\nonumber \\
&=&-2C\left(1+4\alpha \bar{X}_\pm(a)\right)^{-1}a^{-4}
\label{Y_pm}\,.
\end{eqnarray}
This is the advantage in our description of the basic equations;
(\ref{eq:brane_trace}) and (\ref{eq:brane_tracefree}).  The former gives
the generalized Friedmann equation, while the  latter is an algebraic
equation for
$E^0_{~0}$.

Because we have the Birkhoff-type theorem\cite{Wiltshire},  the bulk 
spacetime is
described by a ``black hole"  solution\cite{GB_BH,GB_BH2}, whose metric is
\begin{eqnarray}
ds^2=-f(r)dt^2+f^{-1}(r)dr^2+r^2d\Sigma_k^2
\,,
\label{BH_sol}
\end{eqnarray}
where
\begin{eqnarray}
f(r)=k+{r^2\over 4\alpha}\left(1\mp\sqrt{1+{4\alpha\Lambda\over
3}+{16\alpha \mu\over 3 r^4}}\right)
\,.
\label{BH_f}
\end{eqnarray}
A mass  of a ``black hole" is given by $M={\Omega_k\mu/\kappa^2_5 }$,
where 
$\Omega_k$ is a volume of  3-dimensional space
with a unit radius.
Calculating the curvature tensor, we find
\begin{eqnarray}
E^0_{~0}=C^{01}_{~~~01}=\pm{\mu \over r^4}
\left(1+{4\alpha\Lambda\over 3}+{16\alpha \mu\over
3 r^4}\right)^{-{3\over 2}}\left(1+{4\alpha\Lambda\over 3}+{16\alpha \mu \over
9r^4}\right)\,.
\label{E_OO2}
\end{eqnarray}

Using  Eq. (\ref{X_pm}), we find that Eq.
(\ref{E_OO}) with Eq. (\ref{Y_pm})
  is the same as Eq. (\ref{E_OO2}) with $\mu=3C/2$ and $r=a$.
Hence an integration constant $C$ corresponds to a mass of a ``black hole"
just as in the RS II model. However ``dark radiation" $E^0_{~0}$ is not 
simply a
radiation term but it depends complicatedly on $a$. The signature
$\pm$ in Eq. (\ref{E_OO2}) corresponds to $\pm$ of $\bar{X}_\pm$.

Here we shortly summarize a global structure of the ``black hole" solution
(\ref{BH_sol}). It has a singularity at
$r=0$  if
$\mu\neq 0$ as
\begin{eqnarray}
{\cal R}_{ABCD}{\cal R}^{ABCD} \approx {4\mu\over \alpha  r^{4}}\,.
\end{eqnarray}
 If $1+{4\over 3}\alpha \Lambda >0$,
the upper sign solution of (\ref{BH_f}) has an event horizon if $k\leq 0$ or
$k=1$ and
$\mu>3\alpha$, while
for the lower sign case, a singularity becomes naked
unless $k=-1$ and $\mu<3\alpha$.
For the case of $1+{4\over 3}\alpha \Lambda <0$, 
$r$ is bounded from above as $r\leq r_{\rm max}$, where
\begin{eqnarray}
r_{\rm max}=\left({16\alpha
\mu \over 3|1+{4\over 3}\alpha\Lambda|}\right)^{1/4}\,.
\end{eqnarray}
Although the equation $f(r)=0$ has a positive root  for some restricted conditions,
there appears another singularity at $r=r_{\rm max}$. The curvature
invariant diverges there as
\begin{eqnarray}
{\cal R}_{ABCD}{\cal R}^{ABCD}
\approx {\mu\over 12\alpha r_{\rm max}\left(r_{\rm max}-r\right)^3}\,.
\end{eqnarray}
Since this singularity is timelike and then naked, the above root does
not mean an event
horizon. There is no regular asymptotic region.

Another important property of this solution is about stability.
The lower branch solution in Eq.~(\ref{BH_f}) turns out to be
unstable\cite{GB_BH}.

With these basic equations, several authors analyzed the dynamics of
the universe\cite{GB_brane_cosmology}.
In this paper, assuming the induced gravity model\cite{MMT,Dvali},
we first find a condition for a Minkowski brane.
In the induced gravity model\cite{MMT}, we have
\begin{eqnarray}
\tau^0_{~0}=-(\lambda+\rho)+3\mu^2 X\,,
\end{eqnarray}
where $\lambda$ is a positive tension of a brane, $\rho$ is
the energy density on a brane, and $\mu$ is a mass scale in the induced 
gravity, which
is expected to be the Planck mass.
If we set $\mu=0$, we find the model without induced gravity action
on the brane.
In the Minkowski brane, $X=Y=0$ and $\rho=0$.
 From these conditions with Eqs. (\ref{junc_cond2}) and (\ref{Y_pm}), we 
show that $C=0$.
The real value condition for $\bar{X}$ requires
\begin{eqnarray}
1+{4\over 3}\alpha\Lambda>0 \,.
\label{cond:Lambda}
\end{eqnarray}
Inserting the conditions for the Minkowski brane with $C=0$ into Eq.
(\ref{generalized_Friedmann}),
we find
\begin{eqnarray}
\alpha\kappa_5^4\lambda^2=1-4\alpha\Lambda\mp
\left(1+{4\over 3}\alpha\Lambda\right)^{3/2}.
\label{zero_Lambda}
\end{eqnarray}
This is a tuning condition for zero cosmological constant on the brane.
For the upper branch, when we take a limit of $\alpha\rightarrow 0$, we recover
$\Lambda+\kappa_5^4\lambda^2/6=0$, which is the fine-tuning condition
for the RS II model.
Note that such a limit does not exist for the lower branch, although we have
the Minkowski brane in this branch.
The condition (\ref{cond:Lambda}) gives the possible range for $\lambda$, 
that is,
\begin{eqnarray}
&& 0 \leq \alpha \kappa_5^4 \lambda^2 < 4 ~~~~{\rm for ~~the ~~upper 
~~branch}\,,
\nonumber \\
&& 2 \leq \alpha \kappa_5^4 \lambda^2 < 4 ~~~~{\rm for ~~the ~~lower 
~~branch}\,.
\end{eqnarray}
A de Sitter brane (or anti de Sitter brane) is obtained if $\lambda$ is
larger (or smaller) than that given by Eq.
(\ref{zero_Lambda}). 

Finally, we show an asymptotic
Friedmann equation, by perturbing the Minkowski brane spacetime. Here we
do not impose the tuning condition  (\ref{zero_Lambda}).
Setting $X$, $\rho$ and $C/a^4$ as small variables and expanding Eq.
(\ref{generalized_Friedmann}) up to those  first order terms, we find
the conventional Friedmann equation with dark radiation as
\begin{eqnarray}
H^2+{k\over a^2}={\Lambda_4^{(\pm)} \over 3}+{8\pi G_N^{(\pm)}\over 
3}\rho+{{\cal
C}^{(\pm)}\over a^4},
\end{eqnarray}
where
\begin{eqnarray}
\Lambda_4^{(\pm)}&=&
{\alpha\kappa_5^4\lambda^2-1+4\alpha\Lambda\pm \left(1+{4\over
3}\alpha\Lambda\right)^{3/2}
\over 12\alpha\left(1-{4\over 9}\alpha
\Lambda+{1\over 6}\kappa_5^4\lambda\mu^2\right)}
  \,,\\
8\pi G_N^{(\pm)}&=& {\kappa_5^4 \over 6\left(1-{4\over 9}\alpha
\Lambda+{1\over 6}\kappa_5^4\lambda\mu^2\right)}
\left(\lambda-\mu^2\Lambda_4^{(\pm)}\right) \,,\\
{\cal C}^{(\pm)}&=&
{C\over 3\left(1-{4\over 9}\alpha
\Lambda+{1\over 6}\kappa_5^4\lambda\mu^2\right)}
\left(2\pm\sqrt{1+{4\over 3}\alpha\Lambda}+{8\over 
3}\alpha\Lambda_4^{(\pm)}\right) \,.
\end{eqnarray}
Hence in both branches, we have recovered
the conventional Friedmann universe in an asymptotic form, although  the
lower branch is unstable\cite{GB_BH}.
The early stage of the universe may depend on the parameters as
discussed in Ref.~\cite{GB_brane_cosmology}.

If $1+{4\over 3}\alpha\Lambda<0$, the scale factor of the universe cannot
be infinitely large.  There is an upper bound as
$a<a_{\rm max}=r_{\rm max}$.
No Minkowski brane exists.
If the scale factor approaches  this value, the Weyl curvature (\ref{E_OO2})
diverges,
where a singularity appears in a bulk black hole spacetime.
Hence our universe evolves into a singularity although a scale factor is 
finite.
Even if the universe does not approach this singularity, 
the universe 
will get into trouble because it is a naked singularity.

\section{Concluding Remarks}
\label{IV}

We have derived the covariant gravitational equations of a brane world model
  with the Gauss-Bonnet curvature-squared term in a bulk spacetime.
Although the obtained equations are very complicated, any effects from a 
bulk spacetime
to a brane world are described only by the Weyl curvature ($E_{\mu\nu}$).
The basic equations are not given in a closed form because of this term.

Giving the energy-momentum tensor of the brane, which is shown to be conserved,
the extrinsic curvature ($K_{\mu\nu}$) of a brane satisfies a cubic matrix 
equation.
Since it
is not explicitly given by the energy-momentum tensor, we have to solve a
couple of equations for the induced metric and the extrinsic curvature.
If the brane action includes the induced gravity term, which may be expected 
from quantum
effects of matter fields on the brane, we have to replace the 
energy-momentum tensor with
its generalization  just as in Ref.~\cite{MMT}.

We have then applied the present formalism to cosmology.
Assuming the FRW spacetime for a brane world, we
have re-derived the generalized Friedmann equation.
The obtained equation has one integration constant, just as in the RS II model,
which is proportional to mass of a 5-D black hole solution.
Hence the cosmological model has only one unknown parameter.
The system is described in a closed form.

Note that the present approach can be applied not only to a brane model 
with the Gauss-Bonnet
term in arbitrary dimensions but also to that with any Lovelock terms
because of their quasi-linearity.
Another extension is inclusion of a dilaton field. In a realistic string 
theory, we
have a dilaton field which couples to the Gauss-Bonnet term as well.
It will change the dynamics of a brane world too. Such extensions are in 
progress.
Analyzing those models, we hope that some fundamental cosmological problems
such as a big-bang singularity or a cosmological constant will be solved.

\acknowledgments

We would like to thank K. Aoyanagi, N. Deruelle, S. Mizuno and N. Okuyama  for useful
   discussions.
This work was partially supported by the Grant-in-Aid for Scientific
Research  Fund of MEXT (Nos.
14047216, 14540281) and by the Waseda University Grant for Special
Research Projects.



\end{document}